\documentclass{iopart}
\pdfoutput=1

\usepackage{iopams}
\usepackage{amssymb}
\usepackage{epsfig}
\usepackage{color}
\usepackage{dcolumn}
\usepackage{graphicx}
\usepackage{epstopdf}           
\usepackage[utf8]{inputenc}
\usepackage{latexsym}
\usepackage{times}

\newcommand{\scri}{\ensuremath{\mathcal{J}^+}}

\newcommand{\ie}{\textit{i.e.,\ }}

\newcommand{\news}{\ensuremath{\mathcal{N}}}

\definecolor{rgb_blue}{rgb}{0.0,0.0,0.6}
\definecolor{rgb_green}{rgb}{0.0, 0.6, 0.0}
\definecolor{rgb_red}{rgb}{0.7,0.,0.}

\newcommand{\eqref}[1]{Eq.~(\ref{#1})}

\def\be{\begin{equation}}
\def\ee{\end{equation}}
\def\bea{\begin{eqnarray}}
\def\eea{\end{eqnarray}}

\newcommand\T{\rule{0pt}{2.6ex}}
\newcommand\B{\rule[-1.2ex]{0pt}{0pt}}

\begin{document}


\title{Initial data transients in binary black hole evolutions}

\author{Nigel Bishop}
\address{
  Department of Mathematics,
  Rhodes University,
  Grahamstown, 6139
  South Africa
}

\author{Denis Pollney}
\address{
  Departament de F\'isica,
  Universitat de les Illes Balears,
  Palma de Mallorca,
  E-07122, Spain
}

\author{Christian Reisswig}
\address{
  Theoretical Astrophysics Including Relativity,
  California Institute of Technology,
  Pasadena, CA 91125, USA
}

\date{\today}

\begin{abstract}
We describe a method for initializing characteristic evolutions of the
Einstein equations using a linearized solution corresponding to purely
outgoing radiation. This allows for a more consistent application of
the characteristic (null cone) techniques for invariantly determining
the gravitational radiation content of numerical simulations.
In addition, we are able to identify the {\em ingoing} radiation
contained in the characteristic initial data,
as well as in the initial data of the 3+1
simulation. We find that each component leads to a small but long
lasting (several hundred mass scales) transient in the measured outgoing
gravitational waves.
\end{abstract}


\maketitle


\section{Introduction}
\label{sec:introduction}
It is well known in numerical relativity that current practice for the
setting of initial data introduces spurious radiation into the system,
in both the 3+1 and the characteristic approaches. The error in
the initial data leads to an initial burst of spurious ``junk''
radiation that results from solving the constraint equations on a
single hypersurface, without knowing the past history of the radiation
content. Common practice regards the signal as physical only after
it has settled down following the burst arising from the initial data
solution. While it is straightforward to handle the junk radiation
in this way, a more serious issue is whether the spurious radiation
content of the initial data leads to longer-term transients in the
wave signal. This question has been considered before, but previous
work on the long-term effect of the initial data is
limited~\cite{Bishop05,Kelly2007,Sperhake2007,Lovelace2009}.

Characteristic extraction is a method of invariantly measuring the
gravitational wave emission of an isolated source by transporting the
data to null infinity ($\scri$) using a null formulation of the Einstein
equations~\cite{Bishop98b, Babiuc:2005pg, Babiuc:2009,
  Reisswig:2009us, Reisswig:2009rx, Babiuc:2010ze}.  Initial data is
needed on a null cone in the far field region, say $r> 100M$. Previous
work has taken the simplistic approach of setting the null shear
$J=0$ everywhere, although a recent investigation sets $J$ by the
condition that the Newman-Penrose quantity $\psi_0=0$~\cite{Babiuc:2010ze}.
Setting shear-free initial data is not necessarily incorrect
physically -- for example in the Schwarschild geometry in natural
coordinates $J=0$ everywhere, and it is possible to construct a radiating solution
with $J=0$ everywhere at a specific time. However in the generic case,
a radiating solution has $J\ne 0$, and imposing $J=0$ in effect means that
the outgoing radiation implied by the boundary data must be matched at the
initial time by ingoing radiation.
In principle, the spurious incoming content of $J=0$ data
extends out towards infinity, and so could contaminate the entire
evolution. However, previous results comparing characteristic
extraction with conventional finite radius extrapolation indicate that
it is at most a minor effect~\cite{Reisswig:2009us, Reisswig:2009rx}.

Since the characteristic initial data is needed only in the far field
region, linearized theory provides a suitable approximation. That is,
it is possible to construct initial data that, at the linearized
approximation, represents the physical situation of a gravitational
field with purely outgoing radiation produced by sources in a central
region~\cite{Bishop-2005b}. We first solve the case of two equal mass
objects in circular orbit around each other in a Minkowski background
as a model analytic problem.

We then develop and implement
a procedure to calculate characteristic metric data that contains
purely outgoing radiation. This is done within the context of
characteristic extraction, so that initial data on the null cone is
constructed that is compatible with given data on a worldtube
$\Gamma$ at constant radius. We are then able to compare the waveforms
computed by characteristic extraction using as initial data (a) $J=0$,
and (b) the linearized solution.  We find that while the dominant
gravitational wave modes are largely unaffected by the choice of
initial $J$, a small residual difference is visible between the two
approaches, and can take several hundred $M$ to be damped below other
effects.

Any mis-match between the linearized solution and the actual data is an
indication of an ingoing radiation content. Now, {\em on the worldtube
$\Gamma$}, the characteristic metric data is determined entirely by the
3+1 data so that any ingoing radiation can be traced back to an ingoing
radiation content in the conformally flat 3+1 initial data. In this way we
show that the 3+1 initial data contains a component of ingoing radiation
which results in a long-lasting transient.

The plan of the paper is as follows. Sec.~\ref{s-rev} introduces the
notation, and reviews results needed from previous
work. Sec.~\ref{s-2=mS} applies linearized characteristic theory to
calculate metric data for two equal mass non-spinning black holes in
circular orbit around each other.  Sec.~\ref{s-cotm} describes the
method to construct metric data everywhere from data on a
worldtube. Sec.~\ref{s-res} describes our numerical results, which are
obtained within the context of a binary black hole inspiral and merger.
The paper ends with Sec.~\ref{s-conc}, Discussion and Conclusion.

\section{Review of results needed from other work}
\label{s-rev}

\subsection{The Bondi-Sachs metric}
The formalism for the numerical evolution of Einstein's equations, in
null cone  coordinates, is well known
~\cite{Bondi62,Isaacson83,Bishop96,Bishop97b,Gomez01,Bishop99}.
For the sake of completeness, 
we give a summary of those aspects of the formalism that will be used here.
We start with coordinates based upon a family of outgoing null
hypersurfaces.
Let $u$ label these hypersurfaces, $x^A$ $(A=2,3)$ label
the null rays, and $r$ be a surface area coordinate. In the resulting
$x^\alpha=(u,r,x^A)$ coordinates, the metric takes the Bondi-Sachs
form~\cite{Bondi62,Sachs62}
\begin{eqnarray}
 ds^2  &=&  -\left(e^{2\beta}(1 + W_c r) -r^2h_{AB}U^AU^B\right)du^2
\nonumber \\
        & &- 2e^{2\beta}dudr -2r^2 h_{AB}U^Bdudx^A 
        +  r^2h_{AB}dx^Adx^B,
\label{eq:bmet}
\end{eqnarray}
where $h^{AB}h_{BC}=\delta^A_C$ and
$det(h_{AB})=det(q_{AB})$, with $q_{AB}$ a metric representing a unit
2-sphere; $W_c$ is
a normalized variable used in the code, related to the usual Bondi-Sachs
variable $V$ by $V=r+W_c r^2$. As discussed in more detail below, we
represent $q_{AB}$ by means of a complex dyad $q_A$. Then,
for an arbitrary Bondi-Sachs metric,
$h_{AB}$ can be represented by its dyad component
\begin{equation}
J=h_{AB}q^Aq^B/2,
\end{equation}
with the spherically symmetric case characterized by $J=0$.
We also introduce the fields
\begin{equation}
K=\sqrt{1+J \bar{J}},\qquad
U=U^Aq_A,
\end{equation}
as well as the (complex differential) eth operators $\eth$ and $\bar \eth$
~\cite{Gomez97}.

In the Bondi-Sachs framework, Einstein's equations
$R_{\alpha\beta}=8\pi(T_{\alpha\beta} -\frac{1}{2}g_{\alpha\beta}T)$
are classified as: hypersurface equations --
$R_{11},q^AR_{1A},h^{AB}R_{AB}$ -- forming a hierarchical set for
$\beta,U$ and $W_c$; evolution equation $q^Aq^B R_{AB}$ for $J$; and
constraints $R_{0\alpha}$. An evolution problem is normally formulated
in the region of spacetime between a timelike or null worldtube
$\Gamma$ and future null infinity (${\mathcal J}^+$), with (free)
initial data $J$ given on $u=0$, and with boundary data for
$\beta,U,U_{,r},W_c,J$ satisfying the constraints given on the inner
worldtube. We extend the computational grid to \scri by
compactifying the radial coordinate $r$ by means of a transformation
\begin{equation}
r \rightarrow x=\frac{r}{r+r_{\rm wt}}.
\end{equation}
In characteristic coordinates, the Einstein equations
remain regular at ${\mathcal J}^+$ under such a transformation.

The free initial data for $J$ essentially determines the ingoing
radiation content at the beginning of the evolution. For the case of
binary black hole evolutions in a 3+1 formalism, the initial Cauchy
data is determined by solving the Hamiltonian and momentum
constraints, usually under the assumption of conformal
flatness. Compatible null data initial solutions are not known, and so
we must choose an {\em ansatz} for $J$ which is approximately
compatible. Previous work~\cite{Reisswig:2009us, Reisswig:2009rx} has
simply set $J=0$. In Section~\ref{s-2=mS}, we propose a refinement whereby
$J$ is set according to a linearized solution which is determined by
the Cauchy initial data solution.


\subsection{The spin-weighted formalism and the $\eth$ operator}
A complex dyad $q_A$ is a 2-vector whose real and imaginary parts are
unit vectors that are orthogonal to each other. Further, $q_A$ represents
the metric, and has the properties
\begin{equation}
q^A q_A = 0, \qquad q^A \bar q_A = 2,\qquad
q_{AB}=\frac{1}{2}(q_A\bar{q}_B+\bar{q}_A q_B).
\label{eq:diad.norm.def}
\end{equation}
Note that $q_A$ is not unique, up to a unitary factor: if $q_A$ represents
a given 2-metric, then so does $q^\prime_A=e^{i\alpha}q_A$. Thus,
considerations of simplicity are used in deciding the precise form of
dyad to represent a particular 2-metric.

Having defined a dyad, we may construct complex quantities representing
angular tensor components on the sphere, for example $X_1=T_A q^A$,
$X_2=T^{AB} q_A \bar{q}_B$, $X_3=T^{AB}_C \bar{q}_A\bar{q}_B\bar{q}^C$.
Each object has no free (angular) indices, and has associated with it a spin-weight
$s$ defined as the number of $q$ factors less the number of $\bar{q}$
factors in its definition. For example, $s(X_1)=1,s(X_2)=0,s(X_3)=-3$,
and, in general, $s(X)=-s(\bar{X})$.
We define derivative operators $\eth$ and $\bar{\eth}$ acting on
a quantity $V$ with spin-weight $s$
\begin{equation}
\eth V=q^A \partial_A V + s \Upsilon V,\qquad
\bar{\eth} V=\bar{q}^A \partial_A V - s \bar{\Upsilon} V
\end{equation}
where the spin-weights of $\eth V$ and $\bar{\eth} V$ are $s+1$ and $s-1$,
respectively, and where
\begin{equation}
\Upsilon=-\frac{1}{2} q^A\bar{q}^B\nabla_A q_B.
\label{e-G}
\end{equation}
Some commonly used dyad quantities are
\begin{equation}
\begin{array}{rll}
 &\T \B \mbox{Spherical polars} &\mbox{Stereographic} \\
  \hline
\T \B ds^2=& d\theta^2+sin^2\theta d\phi^2& 4(dq^2+dp^2)/(1+q^2+p^2)^2\\
\T \B q^A=&(1,i\sin\theta)& \frac{1}{2}(1+q^2+p^2)(1,i) \\ 
\T \B \Upsilon=&-\cot\theta &q+ip.
\end{array}
\end{equation}

The spin-weights of the quantities used in the Bondi-Sachs metric are
\begin{equation}
  \begin{array}{lll}
    s(W_c)=s(\beta)=0,\qquad & s(J)=2,\qquad & s(\bar{J})=-2, \\
    s(K)=0,           \qquad & s(U)=1,\qquad & s(\bar{U})=-1.
  \end{array}
\end{equation}

We will be using spin-weighted spherical
harmonics~\cite{Newman-Penrose-1966,Goldberg:1967}
${}_s Y_{\ell m}$, where the suffix ${}_s$ denotes the spin-weight,
and in the case $s=0$ the $s$ will be omitted i.e.
$Y_{\ell m}={}_0 Y_{\ell m}$. It is convenient to make use of the formalism
described in~\cite{Zlochower03, Bishop-2005b}, and have basis functions
whose spin-weight 0 components are purely real; following~\cite{Bishop-2005b},
these are denoted as ${}_sZ_{\ell m}$.
Note that the effect of the $\eth$ operator acting on $Z_{\ell m}$ is
\numparts
\begin{eqnarray}
  \eth Z_{\ell m} & = & \sqrt{\ell(\ell+1)}\;{}_1Z_{\ell m}, \\
  \eth^2 Z_{\ell m} & = &\sqrt{(\ell -1)\ell(\ell+1)(\ell+2)}\;{}_2Z_{\ell m}.
\end{eqnarray}
\endnumparts

\subsection{Solutions to the linearized Einstein equations}

Ref.~\cite{Bishop-2005b} (see also~\cite{Reisswig:2006}) obtained solutions
to the linearized Einstein equations in Bondi-Sachs form using the ansatz
\be
F(u,r,x^A)=\Re(f_{\ell,m}(r)\exp(i\nu u))\,{}_sZ_{\ell,m}
\ee
for a metric coefficient $F$ with spin-weight $s$. Here, we need the
results for linearization about a Minkowski background, in which the spacetime is
vacuum everywhere except on a spherical shell at $r=r_0$.
Strictly speaking, we should be performing the linearization about a
Schwarzschild (or even Kerr) background rather than about Minkowski. In the
Kerr case it is not known how to do so, and in the Schwarzschild case the
difference is that, in Eq.~(\ref{e-j0}), the $1/r^3$ term is replaced by a
term whose leading-order behaviour is also $1/r^3$ but which is not
representable analytically.
We consider the
lowest order case $\ell=2$ and in the exterior of the shell (\ie $r>r_0$),
and describe that part of the solution that represents purely outgoing
gavitational radiation.
\numparts
\label{e-lin-22}
\be
\beta_{2,\nu}(r)=b_1\;\;\mbox{(constant)}
\label{e-b1}
\ee
\be
j_{2,\nu}(r)=(12 b_1+6i\nu c_1+i\nu^3 c_2)\frac{\sqrt{6}}{9}
+\frac{2\sqrt{6}c_1}{r}+\frac{\sqrt{6}c_2}{3r^3}
\label{e-j0}
\ee
\bea
u_{2,\nu}(r)&=&\sqrt{6}\bigg( \frac{\nu^4 c_2+6\nu^2 c_1-12i\nu b_1}{18}
+\frac{2b_1}{r}+\frac{2c_1}{r^2} \nonumber \\
&&-\frac{2i\nu c_2}{3r^3}-\frac{c_2}{2r^4}\bigg)
\eea
\bea
w_{2,\nu}(r)&=& r^2\frac{12i\nu b_1-6\nu^2c_1-\nu^4c_2}{3} \nonumber \\
&&+r\frac{-6b_1+12i\nu c_1+2i\nu^3c_2}{3}
+2\nu^2c_2\nonumber \\
&&-\frac{2i\nu c_2}{r}-\frac{c_2}{r^2}
\label{e-w0}
\eea
\endnumparts
The solution is determined by setting the constant (real valued)
parameters $b_1$, $c_1$ and $c_2$. The gravitational news corresponding
to this solution is given by
\be
\news=\Re(n_{2,\nu}\exp(i\nu u))\,{}_2Z_{2,m}\;\mbox{ with }
n_{2,\nu}=-i\nu^3 c_2 \frac{\sqrt{6}}{6}
\ee
We will also need the solution in the case $\nu=0,\ell=2$ in the
exterior region $r>r_0$
\numparts
\be
\beta_{2,0}(r)=b_0\;\mbox{ (constant) }
\label{e-b20}
\ee
\be
j_{2,0}(r)=\sqrt{6}\left(\frac{4b_0}{3} +\frac{2c_3}{r} +\frac{2c_4}{r^3}
\right)
\ee
\be
u_{2,0}(r)=\sqrt{6}\left(\frac{2b_0}{r} +\frac{2c_3}{r^3} -\frac{3 c_4}{r^4}
\right)
\label{e-U20}
\ee
\be
w_{2,0}(r)=-2 b_0r - \frac{6c_4}{r^3}.
\label{e-w20}
\ee
\endnumparts
in terms of the additional parameters $b_0$, $c_3$ and $c_4$.

\section{Black hole binaries in circular orbit:
         Solution in the linearized limit}
\label{s-2=mS}

The linearized solution described in the previous section will be used
to set initial data on the null cone. We seek a solution which corresponds
roughly to the source of the gravitational radiation which we will eventually
measure, namely a binary black hole system.
In order to be able to apply the linearized theory, we model each black hole
as having a matter density that is described by a Dirac-$\delta$ function whose
location moves uniformly around a spherical shell. More precisely, the matter
density $\rho$ in the spacetime is
\begin{equation}
\rho=\frac{M}{r_0^2}\delta(r-r_0) \delta(\theta-\pi/2) \left(
\delta(\phi - \nu u) +\delta(\phi -\nu u -\pi) \right),
\end{equation}
with respect to Bondi-Sachs coordinates $(u,r,\theta,\phi)$. The mass
of each black hole is $M$, the circular orbit has radius $r_0$, and
the black holes move with angular velocity $\nu$. We next express $\rho$
in terms of spherical harmonics
\be
\rho=\sum_{\ell,m} \Re\left(\rho_{\ell,m} \exp(|m| i\nu u)\right) Z_{\ell,m},
\ee
and apply the usual procedure, that is multiplication by $Z^*_{\ell,m}$ followed
by integration over the sphere, to determine the coefficients $\rho_{\ell,m}$.
We find that, for $\ell\le 2$, the only nonzero coefficients are
\be
  \begin{array}{ll}
    \rho_{0,0}=\delta(r-r_0)\frac{M}{r_0^2\sqrt{\pi}},\qquad&
    \rho_{2,0}=-\delta(r-r_0)\frac{M}{2 r_0^2}\sqrt{\frac{5}{\pi}}, \\
    \rho_{2,2}=\delta(r-r_0)\frac{M}{2 r_0^2}\sqrt{\frac{15}{\pi}},\; &
    \rho_{2,-2}=-i\delta(r-r_0)\frac{M}{2 r_0^2}\sqrt{\frac{15}{\pi}}.
  \end{array}
\ee

In linearized form, the $R_{11}$ Einstein equation is
\be
\beta_{,r}=2\pi r \rho v_1^2,
\label{e-R11}
\ee
where $v_1$ is the covariant component of velocity in the $r$-direction.
Imposing the gauge condition that the coordinates should be such that, on
the worldline of the origin, the metric takes Minkowski form, it follows
that $\beta=0$ there and consequently at all points within $r<r_0$.
Expanding $\beta$ in terms of spherical harmonics
\be
\beta=\sum_{\ell,m} \Re\left(b_{\ell,m} \exp(|m| i\nu u)\right) Z_{\ell,m},
\ee
and integrating Eq.~(\ref{e-R11}), we find the coefficients $b_{\ell,m}$ for
$r>r_0$,
\begin{equation}
  \begin{array}{ll}
    b_{0,0} = \frac{2 M v_1^2}{r_0}\sqrt{\pi},\; \qquad& 
    b_{2,0} = -\frac{M v_1^2}{r_0}\sqrt{5\pi},\; \\
    b_{2,2} = \frac{M v_1^2}{r_0}\sqrt{15\pi},\; &
    b_{2,-2} = -i\frac{M v_1^2}{r_0}\sqrt{15\pi}.
  \end{array}
\end{equation}

The determination of the remaining metric coefficients depends on the
value of $(\ell,m)$. The case $(0,0)$ is straightforward, and we find
for $r>r_0$
\be
J=0,\quad U=0,\quad
\beta=\frac{M v_1^2}{r_0},\quad W_c=-\frac{4Mv_1^2}{r^2}+\frac{2Mv_1^2}{r r_0}.
\ee

The case $(2,0)$ uses the results for a static shell on a Minkowski background
in~\cite{Bishop-2005b}. We solve the jump conditions across the shell
for the various metric quantities~\footnote{Maple script for this purpose
(\texttt{nu0\_regular\_0.map} with output in \texttt{nu0\_regular\_0.out})
are provided in the supplementary data.}. The result is, in the interior,
\be
b_{2,0}=0,\qquad j_{2,0}(r)=-\frac{4M v_1^2 r^2\sqrt{30\pi}}{15r_0^3},
\ee
and in the exterior
\numparts
\begin{eqnarray}
b_{2,0} & = & -\frac{M v_1^2 \sqrt{5\pi}}{r_0}, \\
j_{2,0}(r) & = & \frac{4M v_1^2 \sqrt{30\pi}}{3r_0}\left(-1+\frac{r_0}{r}
  - \frac{r_0^3}{5r^3}\right).
\end{eqnarray}
\label{e-bj20}
\endnumparts

The cases $(2,\pm 2)$ use the results for a dynamic shell on a
Minkowski background in~\cite{Bishop-2005b}~\footnote{The calculation
is provided in the Maple script
\texttt{regular\_0.map} with the output in \texttt{regular\_0.out}
in the supplementary data.}. The script constructs the general solution
inside and outside the shell $r=r_0$, and uses the constraint equations,
as well as regularity conditions at the origin and at infinity, to
eliminate some of the unknown coefficients. It imposes the jump conditions at
the shell, and finds a unique solution for the remaining unknowns. The result
in the exterior is
\numparts
\bea
  \beta & = & \Re\left(b_{2,2}\exp(2i\nu u)\right)\, Z_{2,2} +
    \Re\left(-i b_{2,2}\exp(2i\nu u)\right)\, Z_{2,-2}  \\
  J & = &\Re\left(j_{2,2}(r)\exp(2i\nu u)\right)\, {}_2Z_{2,2} +
    \Re\left(-i j_{2,2}(r)\exp(2i\nu u)\right)\, {}_2Z_{2,-2}
\label{e-j22}
\eea
\endnumparts
where
\be
b_{2,2}=\frac{M v_1^2}{r_0}\sqrt{15\pi},
\label{e-b0}
\ee
and $j_{2,2}(r)$ takes the form given in Eq.~(\ref{e-j0}). The gravitational
news is
\bea
\news & = &\Re(-i(2\nu)^3 c_2 \frac{\sqrt{6}}{6}\exp(i2\nu u))\,{}_2Z_{2,2}
  \nonumber \\
  &&+\Re(-(2\nu)^3 c_2 \frac{\sqrt{6}}{6}\exp(i2\nu u))\,{}_2Z_{2,-2}.
\label{e-N}
\eea
Although the coefficient $c_2$ is complicated, it can be expressed to
leading order in $r_0 \nu$
\be
c_2= \frac{4b_{2,2}}{5}r_0^3.
\label{e-c1c2}
\ee
It follows that, again to leading order in $r_0\nu$, the news is
\bea
\news &=&M v_1^2 r_0^2\nu^3 16 \sqrt{\frac{2\pi}{5}} \times \nonumber \\
&&\left(\Re(-i\exp(2i\nu u))\,{}_2Z_{2,2}+\Re(-\exp(2i\nu u))
\, {}_2Z_{2,-2}\right)
\eea
from which it is easy to deduce, via the Bondi relation, that the rate of
energy loss of the system is
\be
\frac{dE}{du}=-M^2v_1^4r_0^4\nu^6\frac{2^7}{5}.
\ee
In the limit of a low velocity circular orbit, $v_1=1$, $\nu^2=M/(4r_0^3)$,
the above formula reduces to
\be
\frac{dE}{du}=-\frac{2}{5}\frac{M^5}{r_0^5},
\ee
which is identical to that found from the standard quadrupole
formula~\cite{Misner73}.

\section{Constructing the metric from data on a worldtube}
\label{s-cotm}
In characteristic extraction, the Cauchy evolution provides the characteristic
metric
variables $\beta, J, U$ and $W_c$ on the worldtube $\Gamma$, decomposed into 
spherical harmonics ${}_sY_{\ell, m}$, at every time step. In this section, we
develop a method to find coefficients of the linearized solutions that provide
a fit to the actual numerical data at the worldtube (to linear order
and excluding incoming radiation).
Then we use the linearized
solutions with the coefficients just found to predict $J$ everywhere at some
chosen time $u$, and in this way provide initial data for a numerical
characteristic evolution. We restrict attention
to the dominant modes ${}_sY_{2,2}, {}_sY_{2,-2}, {}_sY_{2,0}$.
The method uses a Fourier decomposition in the time domain, and works
well when the data is approximately sinusoidal, with amplitude and frequency
varying slowly. Accordingly, for the binary black hole computation, the method
is applied over a time domain that excludes both the junk radiation and the
merger.

A metric variable $A$ may be written as
\be
A=a_{Y,2} \,{}_sY_{2,2} + a_{Y,0}\, {}_sY_{2,0} + a_{Y,2}^*\, {}_sY_{2,-2}
\ee
where ${}^*$ denotes the complex conjugate. The relationship between the
coefficients of ${}_sY_{2,2}$ and ${}_sY_{2,-2}$ follows theoretically 
from the requirement that the spin weight 0 metric components must be real;
and further the metric data has been checked to confirm that it does satisfy
the relationship. Transforming to the ${}_sZ_{\ell, m}$ basis, we find
\be
A=a_{Z,2} \,{}_sZ_{2,2} +a_{Z,-2}\, {}_sZ_{2,-2}
+a_{Z,0}\,{}_sZ_{2,0},
\ee
where
\be
a_{Z,2} =
\sqrt{2}\Re(a_{Y,2}),\quad a_{Z,-2} =- \sqrt{2} \Im(a_{Y,2}),\quad
a_{Z,0} = a_{Y,0}
\label{e-A}
\ee
so that the metric data on $\Gamma$ can be re-expressed as coefficients of
${}_sZ_{2,2}$, ${}_sZ_{2,-2}$ and ${}_sZ_{2,0}$ at discrete time values.
Although the data is oscillatory in time, it is not at constant frequency
but is a superposition of multiple solutions with different frequencies.
The linearized solutions behave as $e^{i\nu u}$ for fixed $\nu$, so for the
theory to be applicable the next step is to decompose the metric data into
a superposition of constant frequency components. This is achieved by
making a discrete fast Fourier transform of each metric coefficient
\be
a_{Z,2,k}=\sum_{j=1}^L a_{Z,2}(u_j)\exp\left(\frac{-2\pi i(k-1)(j-1)}{L}\right)
\label{e-aZ2}
\ee
where there are $L$ data points over the time interval $(u_1,u_L)$. The
frequency $\nu$ is related to $k$ by
\be
\nu=\frac{2\pi (L-1)(k-1)}{(u_L-u_1)L}.
\ee
We found that $J$ at $\scri$ from the linearized solutions provides a
smoother fit to the actual data if high frequencies are eliminated
(compare Sec.~\ref{s-res}, Fig.~\ref{f-J22scri}),
and so we undertake further processing only for
$k\le L_1$ (with $1<L_1\ll L$); the setting of the Fourier coefficients for
$k>L_1$ is described later.

In the case $k> 1$, for each value of $k$
Eqs. (\ref{e-b1}) to (\ref{e-w0}) evaluated at the worldtube are four equations
for the three unknowns $b_{1,k}, c_{1,k}, c_{2,k}$. Such an over-determined system
can be tackled by a least-squares-fit algorithm, or alternatively by ignoring
one of the equations so making the system uniquely determined. We found that
the reconstructed linearized solution gave a better fit to the actual data at
$\scri$ in the case that Eq.~(\ref{e-w0}) for $W_c$ was ignored. This then means
that a comparison between the actual and reconstructed data for $W_c$ at the
worldtube provides an indication of the error, which is expected because of
(a) incoming radiation in the initial Cauchy data, (b) Fourier transform effects,
and (c) other effects. We discuss item (b) at the end of this
section, and items (a) and (c) in the next section.

In the case $k=1$, $\nu=0$, and Eqs.~(\ref{e-b20}) to (\ref{e-w20}) are four equations
for the constants $b_0, c_3, c_4$. We solved four equations for three unknowns using
a least-squares-fit algorithm, because this approach led to the reconstructed data
having a better fit to the actual data than in the case that Eq.~(\ref{e-w20})
was ignored.

In this way, for a given spherical harmonic say $Z_{2,2}$, we obtain
values for the constants of the frequencies represented by $k=1\cdots
L_1$. Now, our purpose is to use the worldtube data to estimate $J$
off the worldtube.  From Eq.~(\ref{e-j0}) we can write \be
j_{2,k}(r)=d_{0,k}+\frac{d_{1,k}}{r}+\frac{d_{2,k}}{r^3} \ee where for
$2\le k\le L_1$ \numparts \bea
d_{0,k}&=&\frac{\sqrt{6}}{9}\left(12b_{1,k}+6i\nu
c_{1,k}+i\nu^3c_{2,k}\right), \\ d_{1,k}&=&2\sqrt{6}\,
c_{1,k},\\ d_{2,k}&=&\frac{\sqrt{6}}{3}\, c_{1,k}, \eea \endnumparts
and for $k=1$ \be d_{0,1}=\frac{4\sqrt{6}}{3},\quad
d_{1,1}=d_{2,1}=2\sqrt{6}.  \ee We then apply the inverse discrete
Fourier transform to find \be d_0(u)=\frac{1}{L}\sum_{k=1}^{L}
d_{0,k}\exp \left(\frac{2\pi i (k-1) (u-u_1)(L-1)}{(u_L-u_1)L}
\right), \ee where $d_{0,k}=0$ for $L_1+1\le k\le L-L_1+1$, and \be
d_{0,L-k+1}=d_{0,k+1}^*\;\mbox{ for }k=1\cdots (L_1-1).
\label{e-d0L}
\ee 
Eq.~(\ref{e-d0L}) follows from the condition that $d_0(u)$ (and all other
coefficients in the time domain) are real.
The functions $d_1(u)$ and $d_2(u)$ are found in a similar way, and so
we are able to find the coefficient
of $J(u,r)$ of a given spherical harmonic, say ${}_2Z_{2,2}$. Repeating the
calculation for the other spherical harmonics ${}_2Z_{2,-2}$ and ${}_2Z_{2,0}$
leads to a prediction of $J(u,r,x^A)$ to lowest order $\ell=2$.

The coefficient of ${}_2Z_{2,0}$ is not oscillatory, but can rather be
described as slowly varying. While in principle, such behaviour can be
represented by a Fourier decomposition, we found that a better fit was
obtained by regarding the solution as almost constant and solving
Eqs.~(\ref{e-b20}) to (\ref{e-U20}) at each time step, with
Eq.~(\ref{e-w20}) used as a measure of the error
\footnote{The Matlab scripts \texttt{ft\_wt\_driver.m, FT\_WT.m} and
\texttt{nu0.m} are provided in the supplementary data.}.

We investigated the possibility of errors introduced in the Fourier transform
and inverse transform process by comparing \emph{on the worldtube} the actual and
reconstructed values of $\beta$, $J$ and $U$, because the construction is such
that they should be identical\footnote{Note that $W_c$ is not necessarily
identical, 
since there are may be differences due to nonlinearities and incoming radiation.}. 
The following comparison is for the R100 case as
specified in the next section. We found that there was essentially no
difference between the original and reconstructed data, apart from minor
variations over about the first and last $\pm 30M$ of the time interval (of
total duration $1290M$), presumably caused by the cut-off of high frequencies.
This test was performed for both $L_1=50$ and $L_1=100$ with no
visible difference seen in the graphs, indicating that the precise way in
which high frequencies are removed is not important.

\section{Numerical results}
\label{s-res}

\begin{table}[t]
\begin{center}
\begin{tabular}{lccc}
& Worldtube location & Initial time & Initial data \\
\B Data set        & $R_\Gamma [M]$ &  $u_0 [M]$ &  $J$ \\
\hline
\T J0-R100-u0      & $100$                              & $0$                    & $J=0$ \\
J0-R250-u0      & $250$                              & $0$                    & $J=0$ \\
J0-R100-u450    & $100$                              & $450$                  & $J=0$ \\
J0-R250-u900    & $100$                              & $900$                  & $J=0$ \\
Jlin-R100-u450  & $100$                              & $450$                  & $J=J_{\rm lin}$ \\
Jlin-R250-u900  & $250$                              & $900$                  & $J=J_{\rm lin}$ \\
\end{tabular}
\end{center}
\caption{The characteristic evolutions performed based on the same
  Cauchy evolution of two equal mass non-spinning black holes.}
\label{table:models}
\end{table}

We apply the method outlined in the previous section to
the problem of measuring gravitational waves from binary black hole
simulations. Following our implementation of characteristic
extraction~\cite{Reisswig:2009us, Reisswig:2009rx}, we first evolve a
spacetime with a 3+1 (Cauchy) evolution code, recording metric data on
a world tube, $\Gamma$, of fixed coordinate radius. This data is
subsequently used as inner boundary data for a null-cone evolution of
the Einstein equations, which transports the data to $\scri$, where
the gravitational waves are measured. The linearized solution allows
us to specify data for $J$ on the initial null cone which is compatible
(to the linear level) with the data on $\Gamma$ and thus, importantly, to the 
Cauchy 3+1 spacetime.

As a fiducial test case, we return to the well-studied model of an
8-orbit binary system with equal mass non-spinning black holes
carried out in~\cite{Reisswig:2009us, Reisswig:2009rx}. For the Cauchy
evolution, we use the \texttt{Llama} multipatch code described
in~\cite{Pollney:2009ut, Pollney:2009yz}.  We output metric data on
two worldtubes located at $R_\Gamma=100M$ and $R_\Gamma=250M$ that are
used as inner boundary data for a subsequent characteristic evolution.
The waveforms at $\scri$ should be independent
of the worldtube location. Thus, evolutions from 
different worldtube locations help us validate our results.
Table~\ref{table:models}, summarizes the various
characteristic evolutions that we have performed, all of which are
based on the same Cauchy data, but with different characteristic
initial data, $J$, and starting points in Bondi time, $u_0$.

The first approach follows the original prescription laid out
in~\cite{Reisswig:2009us, Reisswig:2009rx}. The characteristic
evolution is started coincident with the first available Cauchy data,
at coordinate time $t_0=u_0=0$.  The characteristic variable $J$ is
initialized by the shear-free solution $J=0$. Since we are beginning
the characteristic evolution from the initial Cauchy slice, these
models include the spurious junk radiation contained in the
conformally flat constraint solution. Models J0-R100-u0 and J0-R250-u0
listed in Table~\ref{table:models} follow this prescription, using
data from the worldtubes at $R_\Gamma=100M$ and $R_\Gamma=250M$,
respectively.

\begin{figure}
\includegraphics[scale=0.7]{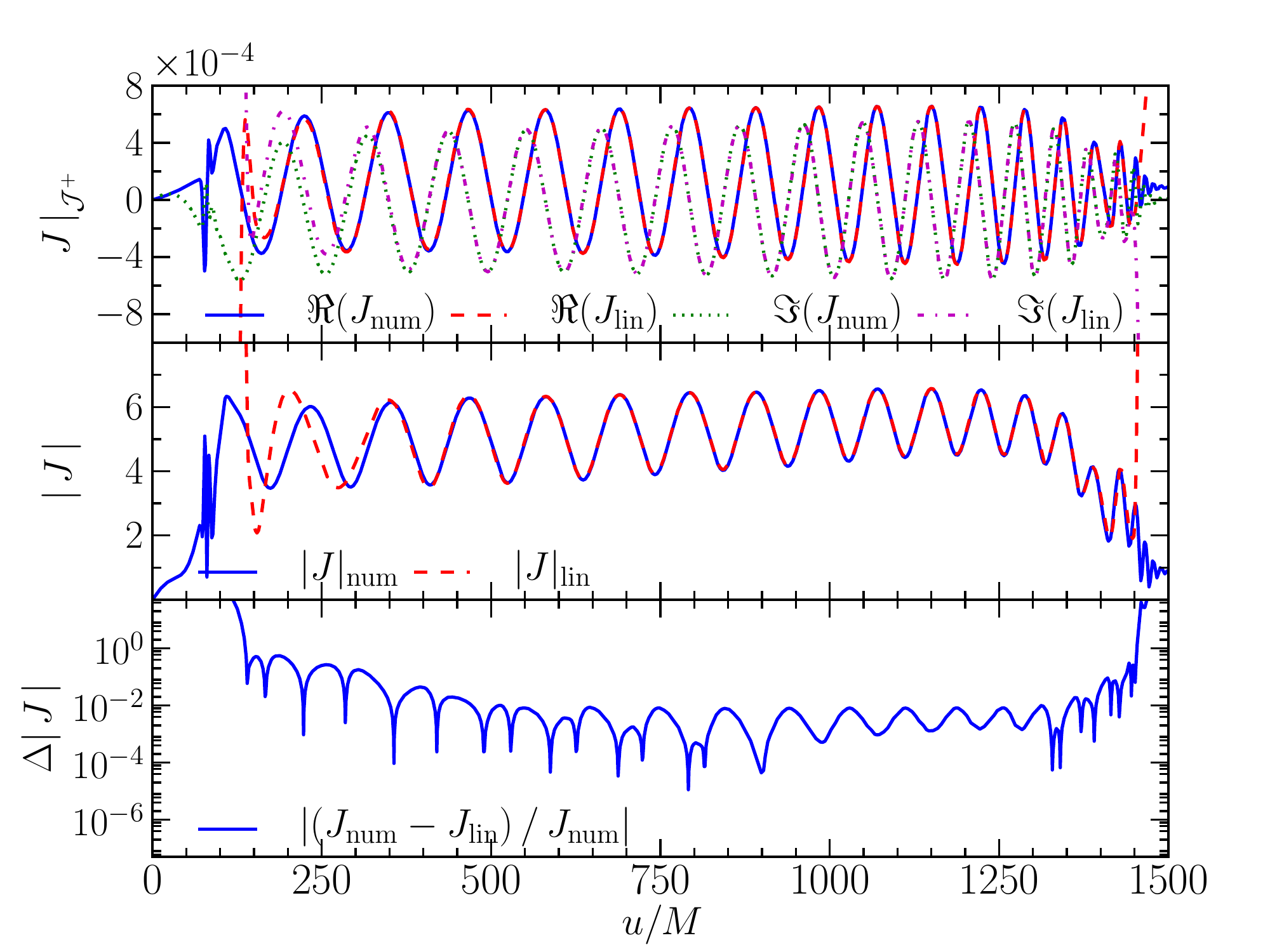}
\caption{Components of the ${}_2Y_{2,2}$ mode of $J$ at $\scri$
  estimated from boundary data at $R_\Gamma=100M$ once using
  linearized solutions, and once using the full non-linear
  characteristic evolution J0-R100-u0. The numerical solution is
  denoted by $J_\mathrm{num}$, while the reconstructed linear solution
  is $J_\mathrm{lin}$. The linearized solution makes use of data
  starting from a time after which the
  initial burst of junk radiation has left the system.  The two
  solutions agree reasonably well only after a time $t_2$
  Eq.~(\ref{eq:incoming-time}), a time after which the incoming
  radiation content of the Cauchy initial data has essentially settled
  to zero.}
\label{f-J22scri}
\end{figure}

It is interesting to compare the results of the fully nonlinear
characteristic Einstein evolution with the corresponding linearized
solution. Fig.~\ref{f-J22scri} plots the $(\ell,m)=(2,2)$ spherical
harmonic modes of $J_\mathrm{num}$, computed by the null Einstein
evolution J0-R100-u0. The linearized solution $J_\mathrm{lin}$ is
computed using linearly reconstructed worldtube data
according to the prescription in Sections~\ref{s-2=mS}
and~\ref{s-cotm}. We compute the linearized solution from the boundary data
at $\Gamma$ after the initial
data junk radiation has passed the worldtube radius $R_\Gamma$, at
around $u=150M$, when the system has settled to the expected binary
black hole inspiral pattern compatible with the solution
construction. The upper panel of Fig.~\ref{f-J22scri} plots the
real and imaginary parts of $J$, evaluated at $\scri$. The center panel
plots the amplitudes of $J$, while the bottom panel shows the relative
difference between the linearly estimated $J_\mathrm{lin}$ and the fully
relativistic result $J_\mathrm{num}$ (model J0-R100-u0).
The linearized $J_{\rm lin}$ and numerically evolved
$J_{\rm num}$ differ initially, but eventually, after around $u=450M$,
differ by less that $1\%$, which remains consistent for the bulk of
the time.

\begin{figure}
\includegraphics[scale=0.7]{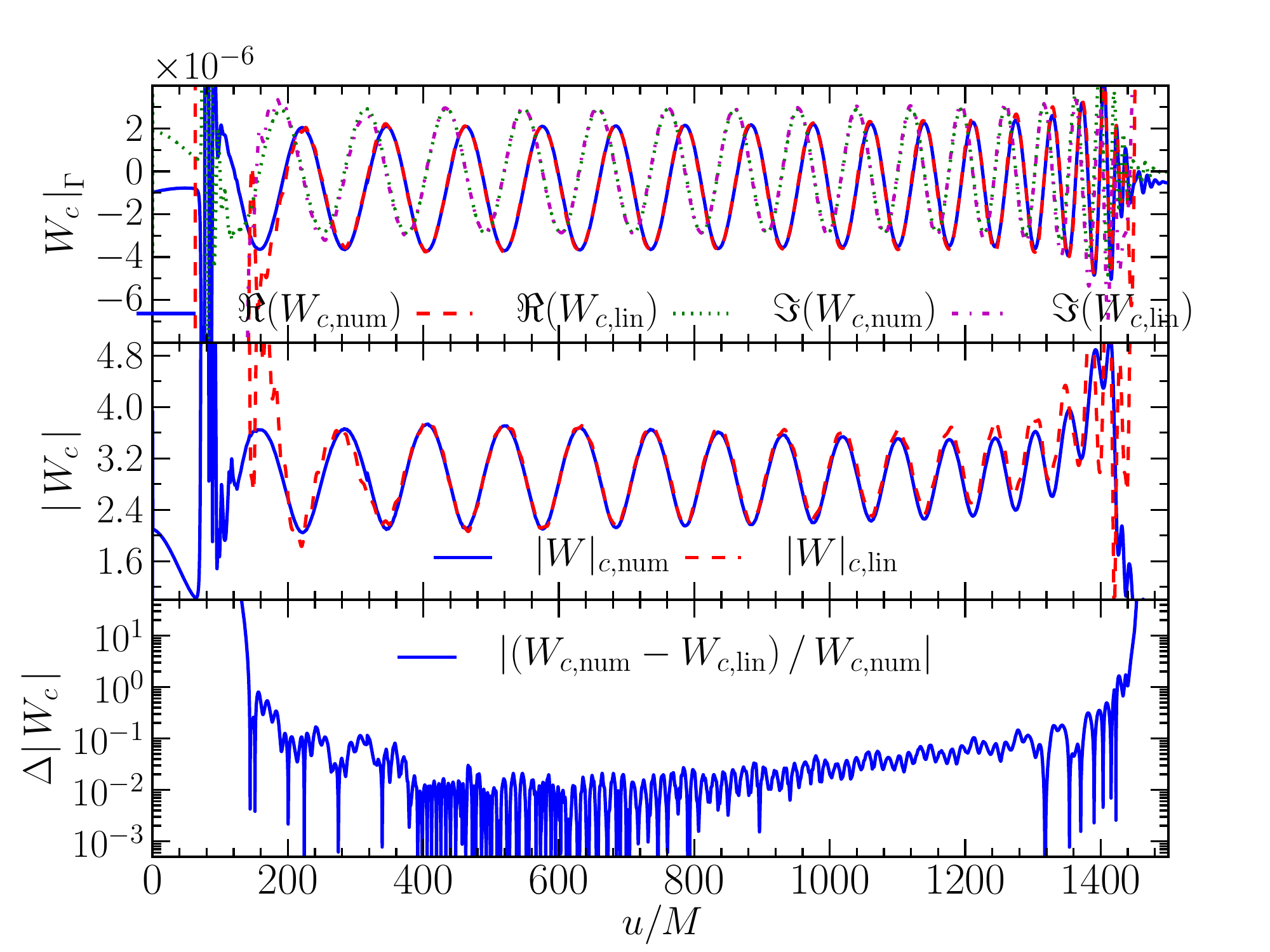}
\caption{
  Components of the ${}_2Y_{2,2}$ mode of $W_c$ at the worldtube
  $R_\Gamma=100M$ once using the linearly reconstructed data, and once
  using the original data as obtained from the Cauchy data. The
  numerically obtained Cauchy data is denoted by $W_{c,\mathrm{num}}$,
  while the reconstructed data is $W_{c,\mathrm{lin}}$. The reconstructed
  data makes use of the original Cauchy data starting from a time
  after which the initial burst of junk
  radiation has left the system.  The numerical and reconstructed data
  agree reasonably well only after a time $t_2$
  Eq.~(\ref{eq:incoming-time}), a time after which the incoming
  radiation content of the Cauchy initial data has essentially settled
  to zero.}
\label{f-W22scri}
\end{figure}

We first discuss whether the difference between the solutions can be due
to effects 
other than ingoing radiation; such effects could include (a) nonlinearity, or
(b) the linearization background being Minkowski rather than Schwarzschild.
Looking at Figs.~\ref{f-J22scri} and \ref{f-W22scri}, we see that to lowest
order the metric components are slowly varying sinusoidal functions; this
statement also applies to the other metric components (graphs not shown).
We would therefore expect that the magnitude of effects (a) and (b) would
be roughly constant, if not with some increase at later times as merger is
approached. Indeed for nonlinear effects we can look at the Einstein equation
for $R_{11}$, which is $\beta_{,r}=0$ to linear order, with the actual value
being an indication of the magnitude of nonlinearity, and we find that this
quantity does slowly increase with time. However, Figs.~\ref{f-J22scri} and
\ref{f-W22scri} show that the difference between the
solutions, until about $450M$, {\em is getting smaller}. 
This indicates that the effect is 
independent of non-linearities and present already at the linear level.
The linearized solution assumes that the radiation is \emph{purely outgoing},
whereas the actual data may contain incoming modes
originating in either the characteristic or Cauchy initial data -- both
options being possible since $J$ at $\scri$ is influenced by both data sets.
Thus, the explanation for Fig.~\ref{f-J22scri} is that
it reflects the slow
decay of the effect of \emph{incoming modes} in the initial data, until saturated
by other factors (such as non-linear effects).
 
A similar effect is seen in the characteristic variable $W_c$, related
to the Newtonian potential, plotted in Fig.~\ref{f-W22scri}.
In this case, however, there is
clarity about the source of the incoming radiation: it must be in the
Cauchy data. This is because in characteristic extraction, the
characteristic metric at the worldtube is determined entirely by the
Cauchy data. Again, the lower panel shows
an approximately exponential decay in the differences, until
around $u=400M$. The residual steady state differences
result from 
other effects, which gradually increase with the
strength of the gravitational radiation towards the binary merger.

The findings above indicate that a physically expected purely outgoing 
inspiral radiation pattern is only present after some time 
\begin{equation}
  u > u_{\rm incoming}\,,
  \label{eq:incoming-time}
\end{equation} 
where $u_{\rm incoming}$ is the length of a time interval until the
incoming radiation content of the Cauchy initial data has settled to a
negligible amount at the given worldtube location.  Hence, in order to
construct physically meaningful and consistent initial data via the
outgoing linearized solution, it is optimal to begin the solution at a
time $u_0>u_{\rm incoming}$ after which both the junk \emph{and}
incoming radiation content of the initial data have subsided.
The results of Figs.~\ref{f-J22scri} and~\ref{f-W22scri} suggest that
the linearized solution provides a reasonably good approximation to
the data $u\approx 450M$, at which time the system has settled into
an outgoing radiative solution.

For instance, in model Jlin-R100-u450, the exact worldtube location is
$R_\Gamma=100.8492$.  Allowing $50M$ for the visible outgoing junk
radiation to pass, we set the time range over which we use
the boundary data for linearized solution construction to
$(u_0,u_f)=(150.192, 1439.856)$, which includes the inspiral, but not
the merger and ringdown.  The time increment in is
given by $du=0.144$, thus comprising 8967 data points. Referring
to Eq.~(\ref{e-aZ2}), we have $L=8957$, $L_1=100$.
In order to determine $J_{\rm lin}$ at the initial time $u_0$, we
use the general form of the linearized solution, \eqref{e-lin-22}:
\begin{eqnarray}
  J&=&\left(e_0+\frac{e_1}{r}+\frac{e_2}{r^3}\right)\,{}_2Y_{2,2}
    +\left(e_0+\frac{e_1}{r}+\frac{e_2}{r^3}\right)^*\,{}_2Y_{2,-2}
  \nonumber
\\ &&+\left(e_3+\frac{e_4}{r}+\frac{e_5}{r^3}\right)\,{}_2Y_{2,0}
  \label{e-Jlin}
\end{eqnarray}
The coefficients are determined by comparing with the worldtube
data. We perform a Fourier transform on the numerically determined
worldtube variables over the time interval $(u_0,u_f)$. 
The spectrum determines the constants $b_1$,
$c_1$ and $c_2$ of  Eqs.~\ref{e-lin-22} at each fixed frequency $\nu$.
These values are transformed back to the time domain, and evaluated
at $t=450M$ in order to determine the coefficients of \eqref{e-Jlin}:
\begin{equation}
  \begin{array}{ll}
    e_0 = (4.5217 + 3.7702 i)\times 10^{-4},\quad & 
    e_1 = -0.04578 - 0.17159 i, \\
    e_2 = 12.582 + 42.500 i, &
    e_3 = -1.4788\times 10^{-4}, \\
    e_4 = 0.020365, &
    e_5 = -35.563.
  \end{array}
  \label{e-450}
\end{equation}

\begin{figure}
  \includegraphics[scale=0.7]{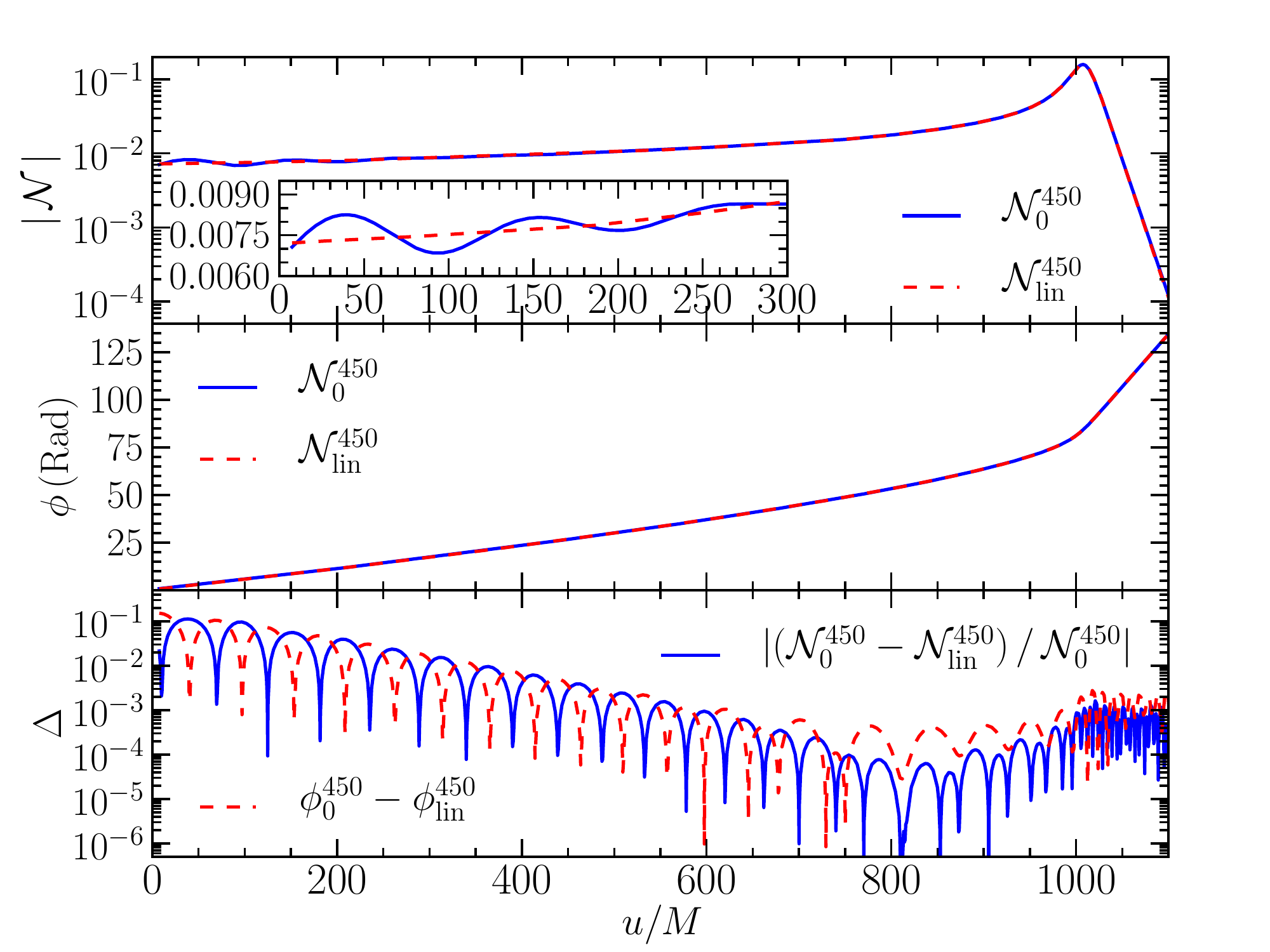}
  \caption{Time domain differences between the $\news_0^{450}$ and
    $\news_\mathrm{lin}^{450}$ waveforms of models J0-R100-u450 and Jlin-R100-u450, 
    computed from worldtube location
    $R_\Gamma=100M$ for which the characteristic runs have been initialized
    by $J=0$ and linearized solutions at a time $u_0=450M$, respectively. 
    The top panel plots the wave amplitude, the middle shows the phase, and the
    lower plots the differences between the two solutions. The
    $\news_0^450$ data shows notable oscillations in amplitude at early
    times (inset), which decay exponentially with time. The waveforms
    have been aligned at the amplitude peak.}
\label{f-diff_amp_vs_time}
\end{figure}

A goal of this paper is to investigate, within the context of
characteristic extraction, the effect of the initial data on the
calculation of the gravitational news.  To this end, we compare
waveforms at $\scri$ from two characteristic evolutions based on the
same Cauchy boundary data, but different initial data constructions:
Jlin-R100-u450 and J0-R100-u450. Both evolutions use boundary data
from $R_\Gamma=100M$ and begin at the initial time $u_0=t_2=450M$,
which was determined above to be a point where the linearized solution
is well-matched to the nonlinear evolution.  The model Jlin-R100-u450
uses the initial data determined by the linearized solutions,
Eqs.~(\ref{e-Jlin}, \ref{e-450}). In contrast, J0-R100-u450 simply sets $J=0$,
corresponding to the original prescription of~\cite{Reisswig:2009us,
  Reisswig:2009rx}.  Fig.~\ref{f-diff_amp_vs_time} plots the Bondi
news at $\scri$ as computed from both evolutions, denoted by
$\news^{450}_{\rm lin}$ and $\news^{450}_0$ for the linearized and
$J=0$ initial data runs, respectively.  Whereas the phase of $\news$
shows very little difference between the runs (middle panel), the
amplitude shows visible oscillations for the $\news^{450}_0$ evolution
(upper panel and inset). The waveforms agree to within $1\%$ only
after a time $u=400M$ (which must be added to the $u_0=450M$ starting
point of the simulation). Therefore, the influence of the $J=0$ {\em ansatz} for
the characteristic initial data has a notable impact over an extended
time.

We note, however, that the original prescription of
\cite{Reisswig:2009us, Reisswig:2009rx} used characteristic initial
data $J=0$ at the initial Cauchy time $u_0=t_0=0$ (that is, including
the junk radiation). At that time, the shear-free approximation
$J=0$, for the characteristic initial data is compatible with the
Cauchy initial data solution.

\begin{figure}
  \includegraphics[scale=0.7]{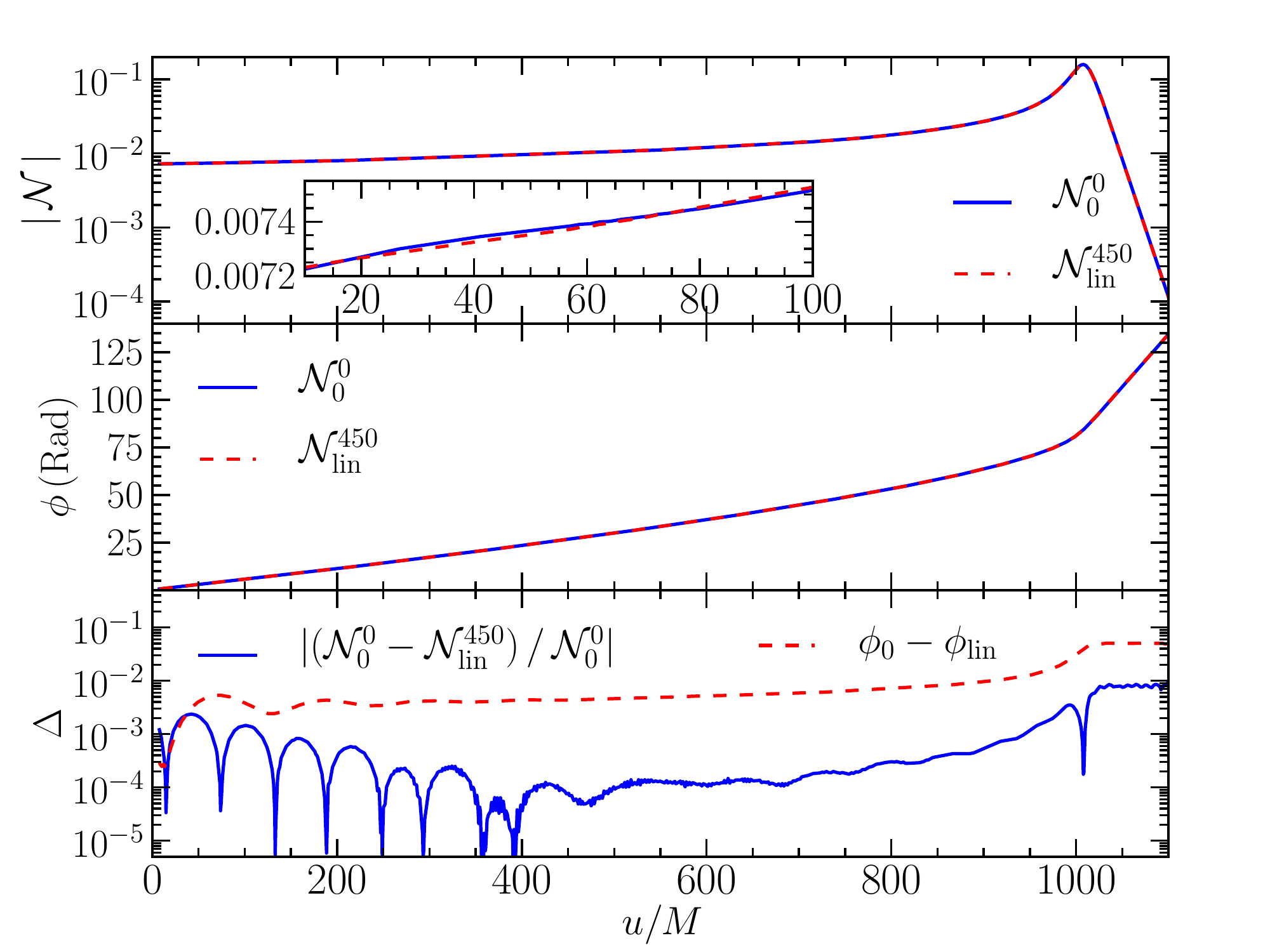}
  \caption{Time domain differences between the $\news_0^0$ (starting off
    at $u_0=t_0=0$) and $\news_\mathrm{lin}^{450}$ (starting off
    at $u_0=450$) waveforms of models J0-R100-u0 and Jlin-R100-u450, computed from
    worldtube location $R_\Gamma=100M$ for which the characteristic runs
    have been initialized by $J=0$ and linearized solutions
    respectively. The top panel plots the wave amplitude, the middle
    shows the phase, and the lower plots the differences between the
    two solutions. The $\news_0^0$ data shows slight oscillations in
    amplitude at early times (inset), which decay exponentially with
    time. The waveforms have been aligned at the amplitude peak.}
\label{f-diff_amp_vs_time-original}
\end{figure}

Hence, we also compare the waveforms of model
Jlin-R100-u450 against those of model J0-R100-u0, where the latter
model uses $J=0$ initial data at time $u_0=0$.  The results are plotted
in Fig.~\ref{f-diff_amp_vs_time-original} with the J0-R100-u0 results
labeled by $\news^0_0$.  We still observe an oscillation in the
amplitude in $\news_0$, though it is drastically reduced compared to
the $\news^{450}_0$ of model J0-R100-u450 shown in
Fig.~\ref{f-diff_amp_vs_time}. The relative errors in amplitude,
plotted in the bottom panel of Fig.~\ref{f-diff_amp_vs_time-original},
are well below $1\%$ over the entire evolution, and the total
dephasing is smaller than $\Delta \phi = 0.04 \rm{rad}$.  We note that
the differences are larger than the systematic error reported in
\cite{Reisswig:2009us, Reisswig:2009rx}.  In that work, the error
estimate refers to the difference between evolutions starting from the
same initial data but different worldtube locations, at a fixed
resolution (though the results converge to the same waveform as the
resolution is increased). Our results indicate that at typical current
resolutions, the choice of characteristic initial data has a larger
influence on the simulation error than the worldtube location.  The
initial burst of junk as well as spurious incoming transients can
alter the measured waveforms by an amount that is of the order of the
discretization error of current numerical relativity codes
(e.g.~\cite{Hannam:2009hh_pbl} and references therein), over a period of several hundred $M$.

The numerical tests described so far were with the
extraction radius $R_\Gamma=100M$. All these runs were repeated
with the extraction radius re-set to $R_\Gamma=250M$. The
results are qualitatively similar to the $R_\Gamma=100M$ case, and
the details are not presented here. One interesting feature was the
behaviour of $W_c$ at the worldtube, \ie the analogy to
Fig.~\ref{f-W22scri}: it took until about $900M$ until the decay in
the difference between the linearized and actual data was saturated.
This is surprising since one would expect that the radiation that passed
$R_\Gamma=100M$ at about $u=450M$ would pass $R_\Gamma=250M$ at $u=600M$.
One possible explanation is that the saturation is due to nonlinear
effects, and since they are somewhat weaker at $R_\Gamma=250M$,
saturation takes longer. Furthermore, the boundary data amplitudes
are one order of magnitude smaller at $R_\Gamma=250M$ than those at
$R_\Gamma=100M$, and thus potentially more sensitive to incoming modes.
Further studies would be required to fully understand the nature of these effects.

\section{Discussion and Conclusion}
\label{s-conc}

Characteristic evolutions provide a means of determining radiated
gravitational energy which is free from the ambiguities associated
with local measures, namely non-linear effects in the near-zone, as well as
ambiguities due to gauge and extrapolations to infinity.  The
principal remaining issue has been to specify appropriate initial data
on the null cone, compatible with the data on the worldtube and the 
Cauchy 3+1 spacetime. The
strong correlation between finite radius results and characteristic
extraction, as well the invariance of the results 
on the worldtube location observed in~\cite{Reisswig:2009us, Reisswig:2009rx}
suggests that even the simplest initial data {\em ansatz}, $J=0$, can
provide results which are accurate enough for astrophysical estimates,
provided the Cauchy 3+1 spacetime is essentially free of radiation at the 
initial characteristic time. 
However, in scenarios where strong outgoing radiation is present during the initial characteristic time, $J=0$
data effectively represents incoming radiation, and may alter significantly
the evolution of the wave signal towards $\scri$. 

The gravitational wave solution developed in Sec.~\ref{s-2=mS}
provides initial data $J=J_{\rm lin}$ which is compatible, to the linearized level, with
outgoing radiation from a binary system. As such, it provides a more
physically motivated starting point than the shear-free, $J=0$,
alternative.  Importantly, we find that the evolutions which take
place from either $J=J_{\rm lin}$ initialized at a time $u_0=450M$ when outgoing radiation is present at the worldtube, 
and $J=0$ at the initial time $u_0=t_0=0$ when the Cauchy 3+1 spacetime is conformally flat
are very similar (compare Fig.~\ref{f-diff_amp_vs_time-original}), and indeed very
similar to the purely 3+1 result which can be obtained by polynomial
extrapolating finite radius measurements. That is, for
simple choices of initial $J$, the physical conclusions are not
altered dramatically, provided the Cauchy 3+1 spacetime does not contain strong amounts of radiation
at the worldtube location during the initial characteristic time $u_0$.

On the other hand, we have demonstrated that the choice of characteristic
initial data does result in a small but measurable difference, which
decays at a slow exponential rate over a time period of several hundred
$M$. We see this transient in the graphs of gravitational news exhibited
in Fig.~\ref{f-diff_amp_vs_time}. Since the linearized initial data contains
only an outgoing mode, we conclude that the shear-free characteristic initial data contains
{\em incoming} radiation. While this is expected, it is interesting that
it takes so much time for the effect to decay away.

This study was designed to assess the long-term effect of characteristic
initial data, but it also provides information about the incoming
radiation in 3+1 initial data. The point is that the characteristic
data on the worldtube is determined entirely by the 3+1 data, and the
extent to which this data does not fit the linearized solution is a
measure of its incoming radiation content. By construction, the quantities
$\beta, U$ and $J$ in the linearized solution must fit the data, with the
difference in $W_c$ being an indication of incoming radiation in the 3+1
evolution. Since $W_c$ is not a gauge invariant quantity, it is not possible
to make a quantitative statement about the magnitude of the incoming radiation,
but Fig.~\ref{f-W22scri} indicates that it takes until at least $\pm 400M$
until it is possible to neglect the effect of incoming radiation in the region
$r<100M$.

The effect of incoming radiation in both 3+1 and characteristic
initial data decays exponentially, but even so it has a more long-term
impact than the outgoing junk radiation which passes the $r=100M$
worldtube radius by approximately $u=150M$.  The effect of incoming
radiation has not received much attention
in previous literature dealing with the problem of initial data
construction. It has potential significance for the construction of
high-accuracy gravitational-wave templates. Further investigations may
also explain properties of the incoming radiation content, giving
further insight to the peeling property of more complex spacetimes
other than Kerr. In particular, it will be important to determine,
using a gauge invariant measure, the long-term effect of incoming radiation
in 3+1 initial data sets.

The results suggest some promising avenues for future study.  Current
methods (``characteristic extraction'') transport data in one
direction (from the Cauchy to the characteristic code).  Great
efficiencies would be possible if the coupling were also carried out
in the other direction, so that the characteristic evolution would
provide outer boundary data for the Cauchy evolution. The linearized
wave solution provides a simple recipe for isolating ingoing
vs. outgoing modes on the characteristic grid, and may be useful in
designing a method for stably transporting data in both directions
across the world tube interface. 

\ack

The authors would like to thank Sascha Husa and C.~D.~Ott for helpful
discussions.  The authors have enjoyed the hospitality of
Rhodes University, the Max-Planck-Institut f\"ur
Gravitationsphysik, Caltech and Universitat de les Illes Balears during
the course of this work. This work was supported by the National Research
Foundation, South Africa, Bundesministerium f\"ur Bildung und
Forschung, and the National Science Foundation under grant numbers
AST-0855535 and OCI-0905046. CR's travel was supported by C.~D.~Ott.
DP has been supported by grants
CSD2007-00042 and FPA-2007-60220 of the Spanish Ministry of Science.
Computations were performed on the NSF Teragrid (allocation
TG-MCA02N014 and TG-PHY100033), the LONI network
(\texttt{www.loni.org}) under allocation \texttt{loni\_numrel05}, the
Barcelona Supercomputing Center, and on the Caltech compute cluster
``Zwicky'' (NSF MRI award No.\ PHY-0960291).


\section*{References}
\bibliographystyle{unsrt}
\bibliography{references,add_refs}

\begin{thebibliography}{10}

\bibitem{Bishop05}
Nigel~T. Bishop, Roberto G{\'o}mez, Luis Lehner, Manoj Maharaj, and Jeffrey
  Winicour.
\newblock Characteristic initial data for a star orbiting a black hole.
\newblock {\em Phys. Rev. D}, 72:024002, 2005.

\bibitem{Kelly2007}
Bernard~J Kelly, Wolfgang Tichy, Manuela Campanelli, and Bernard~F Whiting.
\newblock Black hole puncture initial data with realistic gravitational wave
  content.
\newblock {\em Phys.Rev. D}, 76:024008, 2007.

\bibitem{Sperhake2007}
Ulrich Sperhake.
\newblock Binary black-hole evolutions of excision and puncture data.
\newblock {\em Phys.Rev. D}, 76:104015, 2007.

\bibitem{Lovelace2009}
Geoffrey Lovelace.
\newblock Reducing spurious gravitational radiation in binary-black-hole
  simulations by using conformally curved initial data.
\newblock {\em Class. Quantum Grav.}, 26:114002, 2009.

\bibitem{Bishop98b}
N.~Bishop, R.~Isaacson, R.~G{\'o}mez, L.~Lehner, B.~Szil{\'a}gyi, and
  J.~Winicour.
\newblock In B.~Iyer and B.~Bhawal, editors, {\em Black {H}oles,
  {G}ravitational {R}adiation and the {U}niverse}, page 393. Kluwer, Dordrecht,
  The Neterlands, 1999.

\bibitem{Babiuc:2005pg}
Maria Babiuc, Bel{\'a} Szil{\'a}gyi, Ian Hawke, and Yosef Zlochower.
\newblock Gravitational wave extraction based on {C}auchy-characteristic
  extraction and characteristic evolution.
\newblock {\em Class. Quantum Grav.}, 22:5089--5108, 2005.

\bibitem{Babiuc:2009}
M.~C. Babiuc, N.~T. Bishop, B.~Szil{\'a}gyi, and Jeffrey Winicour.
\newblock Strategies for the characteristic extraction of gravitational
  waveforms.
\newblock {\em Phys. Rev.}, D79:084011, 2009.

\bibitem{Reisswig:2009us}
C.~Reisswig, N.~T. Bishop, D.~Pollney, and B.~Szilagyi.
\newblock {Unambiguous determination of gravitational waveforms from binary
  black hole mergers}.
\newblock {\em Phys. Rev. Lett.}, 103:221101, 2009.

\bibitem{Reisswig:2009rx}
C.~Reisswig, N.~T. Bishop, D.~Pollney, and B.~Szilagyi.
\newblock {Characteristic extraction in numerical relativity: binary black hole
  merger waveforms at null infinity}.
\newblock {\em Class. Quant. Grav.}, 27:075014, 2010.

\bibitem{Babiuc:2010ze}
M.C. Babiuc, B.~Szilagyi, J.~Winicour, and Y.~Zlochower.
\newblock {A Characteristic Extraction Tool for Gravitational Waveforms}.
\newblock {\em arXiv:1011.4223}, 2010.

\bibitem{Bishop-2005b}
Nigel~T. Bishop.
\newblock Linearized solutions of the {E}instein equations within a
  {B}ondi-{S}achs framework, and implications for boundary conditions in
  numerical simulations.
\newblock {\em Class. Quantum Grav.}, 22(12):2393--2406, 2005.

\bibitem{Bondi62}
H.~Bondi, M.~G.~J. van~der Burg, and A.~W.~K. Metzner.
\newblock Gravitational waves in general relativity {VII}. {W}aves from
  axi-symmetric isolated systems.
\newblock {\em Proc. R. Soc. London}, A269:21--52, 1962.

\bibitem{Isaacson83}
R.~Isaacson, J.~Welling, and Jeffrey Winicour.
\newblock Null cone computation of gravitational radiation.
\newblock {\em J. Math. Phys.}, 24:1824, 1983.

\bibitem{Bishop96}
N.~T. Bishop, R.~G{\'o}mez, L.~Lehner, and J.~Winicour.
\newblock {C}auchy-characteristic extraction in numerical relativity.
\newblock {\em Phys. Rev. D}, 54:6153--6165, 1996.

\bibitem{Bishop97b}
Nigel~T. Bishop, Roberto G{\'o}mez, Luis Lehner, Manoj Maharaj, and Jeffrey
  Winicour.
\newblock High-powered gravitational news.
\newblock {\em Phys. Rev. D}, 56(10):6298--6309, 15 November 1997.

\bibitem{Gomez01}
R.~G{\'o}mez.
\newblock Gravitational waveforms with controlled accuracy.
\newblock {\em Phys. Rev. D}, 64:024007, 2001.
\newblock gr-qc/0103011.

\bibitem{Bishop99}
Nigel~T. Bishop, Roberto G{\'o}mez, Luis Lehner, Manoj Maharaj, and Jeffrey
  Winicour.
\newblock Incorporation of matter into characteristic numerical relativity.
\newblock {\em Phys. Rev. D}, 60:024005, 1999.

\bibitem{Sachs62}
R.K. Sachs.
\newblock Gravitational waves in general relativity {VIII}. {W}aves in
  asymptotically flat space-time.
\newblock {\em Proc. Roy. Soc. London}, A270:103--126, 1962.

\bibitem{Gomez97}
Roberto G{\'o}mez, Luis Lehner, Philippos Papadopoulos, and Jeffrey Winicour.
\newblock The eth formalism in numerical relativity.
\newblock {\em Class. Quantum Grav.}, 14(4):977--990, 1997.

\bibitem{Newman-Penrose-1966}
Ezra~T. Newman and Roger Penrose.
\newblock Note on the {B}ondi-{M}etzner-{S}achs group.
\newblock {\em J. Math. Phys.}, 7(5):863--870, May 1966.

\bibitem{Goldberg:1967}
J.~N. Goldberg, A.~J. MacFarlane, Ezra~T. Newman, F.~Rohrlich, and E.~C.~G.
  Sudarshan.
\newblock Spin-$s$ spherical harmonics and $\eth$.
\newblock {\em J. Math. Phys.}, 8(11):2155--2161, 1967.

\bibitem{Zlochower03}
Y.~Zlochower, R.~G{\'o}mez, S.~Husa, L.~Lehner, and J.~Winicour.
\newblock Mode coupling in the nonlinear response of black holes.
\newblock {\em Phys. Rev. D}, 68:084014, 2003.

\bibitem{Reisswig:2006}
Christian Reisswig, Nigel~T. Bishop, Chi~Wai Lai, Jonathan Thornburg, and
  Bel{\'a} Szil{\'a}gyi.
\newblock Characteristic evolutions in numerical relativity using six angular
  patches.
\newblock {\em Class. Quantum Grav.}, 24:S327--S339, 2007.

\bibitem{Misner73}
Charles~W. Misner, Kip~S. Thorne, and John~A. Wheeler.
\newblock {\em Gravitation}.
\newblock W. H. Freeman, San Francisco, 1973.

\bibitem{Pollney:2009ut}
Denis Pollney, Christian Reisswig, Nils Dorband, Erik Schnetter, and Peter
  Diener.
\newblock {The Asymptotic Falloff of Local Waveform Measurements in Numerical
  Relativity}.
\newblock {\em Phys. Rev.}, D80:121502, 2009.

\bibitem{Pollney:2009yz}
Denis Pollney, Christian Reisswig, Erik Schnetter, Nils Dorband, and Peter
  Diener.
\newblock {High accuracy binary black hole simulations with an extended wave
  zone}.
\newblock {\em arXiv:0910.3803}, 2009.

\bibitem{Hannam:2009hh_pbl}
M.~{Hannam} et~al.
\newblock {Samurai project: Verifying the consistency of black-hole-binary
  waveforms for gravitational-wave detection}.
\newblock {\em Physical Review D}, 79(8):084025, April 2009.

\end{thebibliography}

\end{document}